\documentclass[twocolumn,prd,nofootinbib,aps,prl,floats,floatfix,amsmath,amssymb,longbibliography,secnumarab
ic,superscriptaddress,preprintnumbers]{revtex4-1} %
\usepackage[final]{graphicx}
\usepackage{hyperref}
\usepackage{amsmath}
\usepackage{bbm}
\usepackage{bm}
\usepackage{amsfonts}
\usepackage{amssymb}
\usepackage{latexsym}
\usepackage{graphicx}
\usepackage[english]{babel}
\usepackage{multirow}
\usepackage{float}
\usepackage{url}
\usepackage{slashed}
\usepackage{xcolor} 
\usepackage[utf8]{inputenc}
\usepackage{stmaryrd} 
\usepackage{enumitem}
\usepackage{hyperref}
\usepackage{cleveref}
\usepackage{siunitx}
\usepackage{verbatim}

\newcommand{\be}{\begin{equation}}
\newcommand{\ee}{\end{equation}}
\newcommand{\ba}{\begin{array}}
\newcommand{\ea}{\end{array}}
\newcommand{\bea}{\begin{eqnarray}}
\newcommand{\eea}{\end{eqnarray}}
\newcommand{\sss}{\scriptscriptstyle}
\renewcommand{\d}{\mathrm{d}}

\newcommand{\nn}{\nonumber}

\newcommand{\lsm}{\ell_{\sss SM}}
\newcommand{\lbsm}{\ell_{\sss BSM}}

\newcommand{\besub}{\begin{subequations}}
\newcommand{\eesub}{\end{subequations}}

\begin{document}

\preprint{CERN-TH-2019-057}

\title{Can the ANITA anomalous events be due to new physics?}

\author{James M.\ Cline}
\affiliation{McGill University, Department of Physics, 3600 University St.,
Montr\'eal, QC H3A2T8 Canada}
\author{Christian Gross}
\affiliation{Theoretical Physics Department, CERN, Geneva, Switzerland}
\affiliation{Dipartimento di Fisica dell'Universit{\`a} di Pisa, Pisa, Italy}
\author{Wei Xue}
\affiliation{Theoretical Physics Department, CERN, Geneva, Switzerland}

\begin{abstract}

The ANITA collaboration has observed two ultra-high-energy upgoing air
shower events that cannot originate from Standard Model neutrinos that
have traversed the Earth. Several beyond-the-standard-model
physics scenarios have been proposed as explanations for these events. 
In this paper we present some general arguments making it challenging for new physics to explain the events.
One exceptional class of models that could work is pointed out, in which metastable
dark matter decays to a highly boosted lighter dark matter particle, 
that can interact in the Earth to produce the observed events.

\end{abstract}
\maketitle

\section{Introduction}

High energy neutrino astrophysics could provide a window to new
physics at energies beyond those accessible on Earth, thanks to 
the long propagation distances possible for ultra-high-energy
(UHE) neutrinos, compared to charged particles.  The Auger and IceCube
experiments are sensitive to UHE neutrinos, with several events at PeV
energies now observed by IceCube.  Upper limits on the flux at higher
energies are established, up to energies of $10^{11}$\,GeV.

The cross section for neutrinos to interact with nucleons through deep inelastic scattering 
(DIS)~\cite{Connolly:2011vc} grows with energy, and at ultra high energies above $\sim 10^8$~GeV the Earth becomes opaque to neutrinos.
A promising strategy then is to search for Earth-skimming neutrinos that produce charged leptons
with high probability, that do not lose too much energy before they exit \cite{Feng:2001ue}. The vast majority of such events would be
initiated by $\nu_\tau$ converting to $\tau$ leptons via DIS since
electrons or muons would be absorbed by the ice over much shorter
distances and have a negligible probability to exit compared to
$\tau$'s.

ANITA is a balloon-borne instrument that detects 
polarized radio emission from the electromagnetic component of cosmic
ray showers in the atmosphere, reflected from ice at the South Pole,
or arriving without reflection from events near the horizon. 
Because of phase inversion of the signal upon reflection, the
difference between such events is distinguishable, and the zenith
angle of the $\tau$ whose decay produced the shower can be determined to within $0.3^\circ$.

ANITA has made four flights since 2007, and observed two anomalously 
steep events, henceforth called ANITA anomalous events (AAEs), apparently emerging from the Earth, of energy $E_\nu\sim
0.6$\,EeV on the first~\cite{Gorham:2016zah} and third~\cite{Gorham:2018ydl} flights, corresponding to particles
that traversed chord lengths of 5700 and 7200 km, respectively.\footnote{Recently ref.~\cite{deVries:2019gzs} argued that the observed inverted polarity of the AAEs might in fact be explained from down-going air showers.}
The mean free path of neutrinos at this energy is a few hundred km,
making it exceedingly unlikely that standard model processes could
explain these events \cite{Fox:2018syq}.

Several models beyond the Standard Model (SM) have been suggested as possible
explanations.  These fall into three broad categories: (1) 
SM neutrinos of astrophysical origin could convert to 
beyond-the-standard-model (BSM) particles through interactions in 
the Earth, followed by propagation of the BSM particle until it reconverts
to SM particles that initiate the observed hadronic air shower
\cite{Fox:2018syq,Collins:2018jpg,Chauhan:2018lnq,Anchordoqui:2018ssd}; 
(2) dark matter (DM) that has accumulated within the Earth and 
decays to BSM particles that reconvert to 
SM particles shortly below the Antarctic surface could induce 
the air showers;
or (3) an exotic flux of BSM particles,
such as sterile neutrinos \cite{Cherry:2018rxj,Huang:2018als} incident upon the Earth interact
to produce the observed particles near the Earth's surface.  A
possible source of such flux is the decay of long-lived DM
particles \cite{Dudas:2018npp,Anchordoqui:2018ucj,Heurtier:2019git}.

A generic challenge for models in the first category is that, despite
the increased probability for a BSM particle to traverse the Earth
compared to EeV-energy neutrinos, there is still a large reduction in
efficiency because of the need for this particle to reconvert within a
thin layer near the surface of the Earth before it exits into the
atmosphere.  A major goal of the present work is to carefully quantify
this efficiency and to show that it puts such explanations of the
ANITA anomalous events at odds with null searches by IceCube and 
other experiments such as Auger.
Models in the second category face the problem that the amount
of DM particles that may have accumulated within the Earth cannot be large enough to explain the two AAEs.
On the other hand, third category models have the advantage that
the small efficiency factor can be overcome by assuming a larger flux
of sterile BSM particles, that is relatively less constrained 
than the neutrino flux.
Nevertheless such a large flux of sterile BSM particles can be 
problematic if at the same time active neutrinos or other 
nonsterile particles are produced.
We will argue that previously 
proposed models of this kind are unlikely to be viable, but
demonstrate an example of a model that does work.
The essential ingredient is an ultraheavy DM particle
that decays exclusively to another, much lighter, DM
particle,
whose flux is unconstrained. This boosted decay product then
converts to $\tau$'s within the Earth which in turn induce the
hadronic air showers.

The paper is structured as follows:
In Section~II, we perform a model-independent analysis of scenarios
where a neutrino scatters into weakly-interacting states that
can traverse the Earth. These states will convert to $\tau$ or directly decay
to hadrons seen by ANITA. Single conversion and cascade decay processes are considered here, but
neither diffuse nor pointlike $\nu_\tau$ fluxes can explain the AAEs.
Constraints on models that attempt to explain the AAEs from dark matter decays either within the Earth or in the galactic halo are discussed in Section III. In either case, we find that it is not possible to achieve
a large enough flux to explain the ANITA events, while remaining
consistent with limits from IceCube and other experiments.
In Section IV we give an example of a class of models that 
can however be consistent,
involving decays of a heavy subdominant component of metastable
DM into the lighter dominant DM particle, that can interact with
matter in the Earth to produce the anomalous events.
We conclude in Section V.

\section{ANITA events from neutrino flux}
\label{sec:neutrino}

In this section, we analyze the possibility that the ANITA events are
consistent with BSM explanations, assuming
that the BSM particles come from ultra-high-energy (UHE) neutrinos
interacting with matter in the Earth.  Other
sources of BSM particles will be considered in following sections.

We assume that the ANITA anomalous events are air showers
initiated by the hadronic decay of energetic $\tau$ leptons, or alternatively by the
decay of BSM particles directly into hadrons.\footnote{The energetic $\tau$'s which 
induce air showers above
the Antarctic surface can also pass through the IceCube detector,
where they could be misidentified as highly energetic muon tracks. 
At most three
such events have been seen~\cite{Fox:2018syq}, while a rough estimate
of the number of events is approximately one order of magnitude larger than the
number of AAEs~\cite{Huang:2018als,Fox:2018syq}. It is however
possible that a detailed analysis, including in particular instrumental
effects, could lead to a lower number of expected events at
IceCube, thus resolving this tension and making $\tau$-induced air
showers viable (cf.~ref.\ \cite{Fox:2018syq}). Hence we discuss both the
case of $\tau$-induced air showers and air showers directly induced by
a BSM particle in this paper.}
The predicted number of such events is the convolution of 
the differential exposure of ANITA to up-going air showers,
\be
d H_\textrm{\tiny AN}={\rm d}A \, {\rm d}\Omega_{d}  \sin\theta_{\rm em} \, 
      {\rm d} t  \, P_{\rm obs} \ , 
\ee      
and the differential diffuse neutrino flux $\phi(E_\nu) \, \d E_\nu$:
\begin{equation}
\mu =
   \int  
      {\rm d} H_\textrm{\tiny AN}\,
      {\rm d} E_\nu  \, \phi ( E_\nu ) \,.
\label{eq:expectmu}
\end{equation}
Here 
\begin{itemize}
\item
$\d A = R_\varoplus^2 \sin\theta_E \d \theta_E  \d \phi_E$ is the
differential area of the Earth observed by ANITA, with $\theta_E$
being the angle between the South Pole and a point on the surface as seen from the center of the Earth, $\phi_E$ being the
longitude, and $R_\varoplus=6371$~km  the Earth's radius;
\item
$\theta_{\rm em}$ is the 
emergence angle of the $\tau$ (respectively the BSM particle that 
causes the hadronic air shower), measured with respect to the 
horizon---see fig.\ \ref{fig:path}; 
\item$P_{\rm obs}$ is the
probability that an UHE $\nu$ that enters the Earth with energy $E_\nu$ leads to a $\tau$ (or possibly some BSM particle) that exits the Earth near the south pole at an emergence angle $\theta_{\rm em}$ with energy $\bar E_\tau$ and initiates a hadronic air shower within
$D=10~{\rm km}$;\footnote{For altitudes larger than $6\,$km the
typical shower is already starting to exit the atmosphere before
reaching its maximum \cite{Gorham:2016zah}.
We choose $D=10~{\rm km}$ in order to be conservative in constraining
BSM scenarios.}
\item
${\rm d}\Omega_d$ is the detection solid angle of ANITA.
It depends on $\bar E_\tau$ and the $\tau$'s decay position, which requires detailed  simulations of
Extensive Air Showers~(EAS)~\cite{Romero-Wolf:2018zxt}.
Here we
neglect this dependence, making the approximation that 
${\rm d}\Omega_d$ can be integrated independently.
\end{itemize}
The live time of the two experiments was  $ 17.25$ days for
ANITA~I \cite{Schoorlemmer:2015afa} and $7$
days for ANITA III~\cite{Gorham:2018ydl}.  ANITA II was not sensitive
to up-going air showers, and data from ANITA IV have not been released so far.

\subsection{Conversion probability}
\label{sec:conversion}

In the following we will compute upper limits on $P_{\rm obs}$ for 
several ``process topologies,'' independent of a specific BSM realization. 
In the SM the process topology is
\[
 \big\langle 0\big\rangle \qquad \nu \rightarrow
\tau \to {\rm hadrons} \,, 
\qquad\qquad\qquad
\qquad\qquad\qquad
\]
which is  understood to include $\tau$ regeneration and energy loss. 
The BSM process will occur in parallel with any new physics processes,
and so its effects should be included alongside the latter
for quantitative results, even if the SM contribution is quite small.

Assuming the presence of BSM physics, several topologies are 
conceivable. In the first, 
the UHE $\nu$ converts to a BSM particle
$X$, which subsequently converts or decays to a $\tau$:   
\[
 \big\langle 1\big\rangle\qquad  \nu \rightarrow X \rightarrow \tau  \to {\rm hadrons}  \, .
\qquad\qquad\qquad\qquad\quad
\] 
A second possibility is to have cascade decays of several BSM particles, 
\[
	 \big\langle 2\big\rangle \qquad \nu \rightarrow X_1
\rightarrow X_2\to \dots \to X_n\ \to \tau \to {\rm hadrons} \, .
\]
This includes  \big\langle 1\big\rangle\ as the special case $n =1$.
As we will show below, for $n>1$ one can  increase $P_{\rm obs}$ somewhat.  
It is also possible
that $X$ decays directly into hadrons that initiate the observed
air showers.  We denote this by
\[
 \big\langle 3\big\rangle \qquad \nu \rightarrow X_1
\rightarrow X_2\to \dots \to X_n\ \to {\rm hadrons} \ .\ \ \  
\quad
\]

%
\begin{figure}
\includegraphics[width=0.9\hsize]{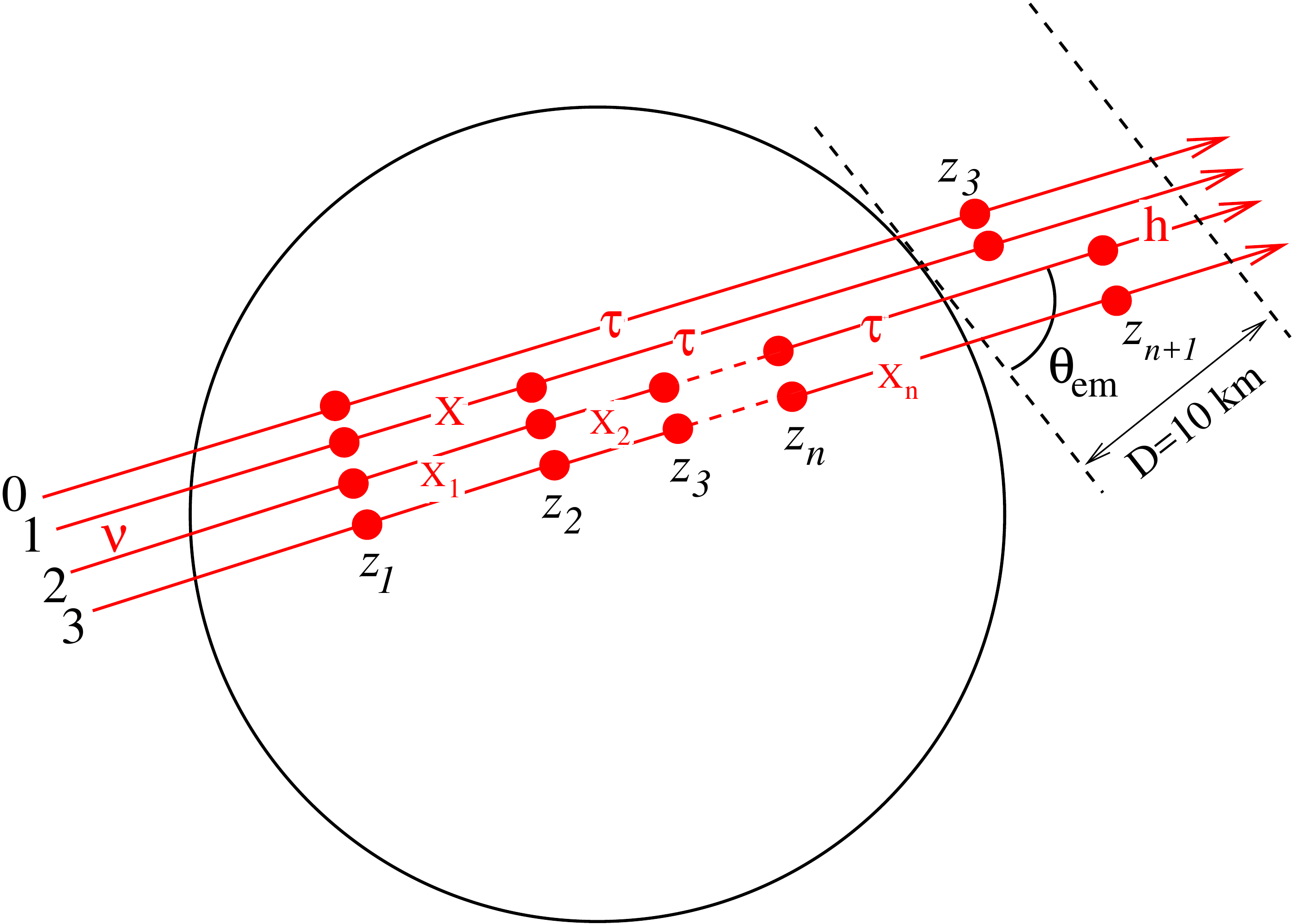}
\caption{Schematic illustration of the processes $0-3$ considered, and the positions
$z_i$ that are integrated over. }
\label{fig:path}
\end{figure}
%

%
%

%
\begin{figure*}
\includegraphics[width=0.485\hsize]{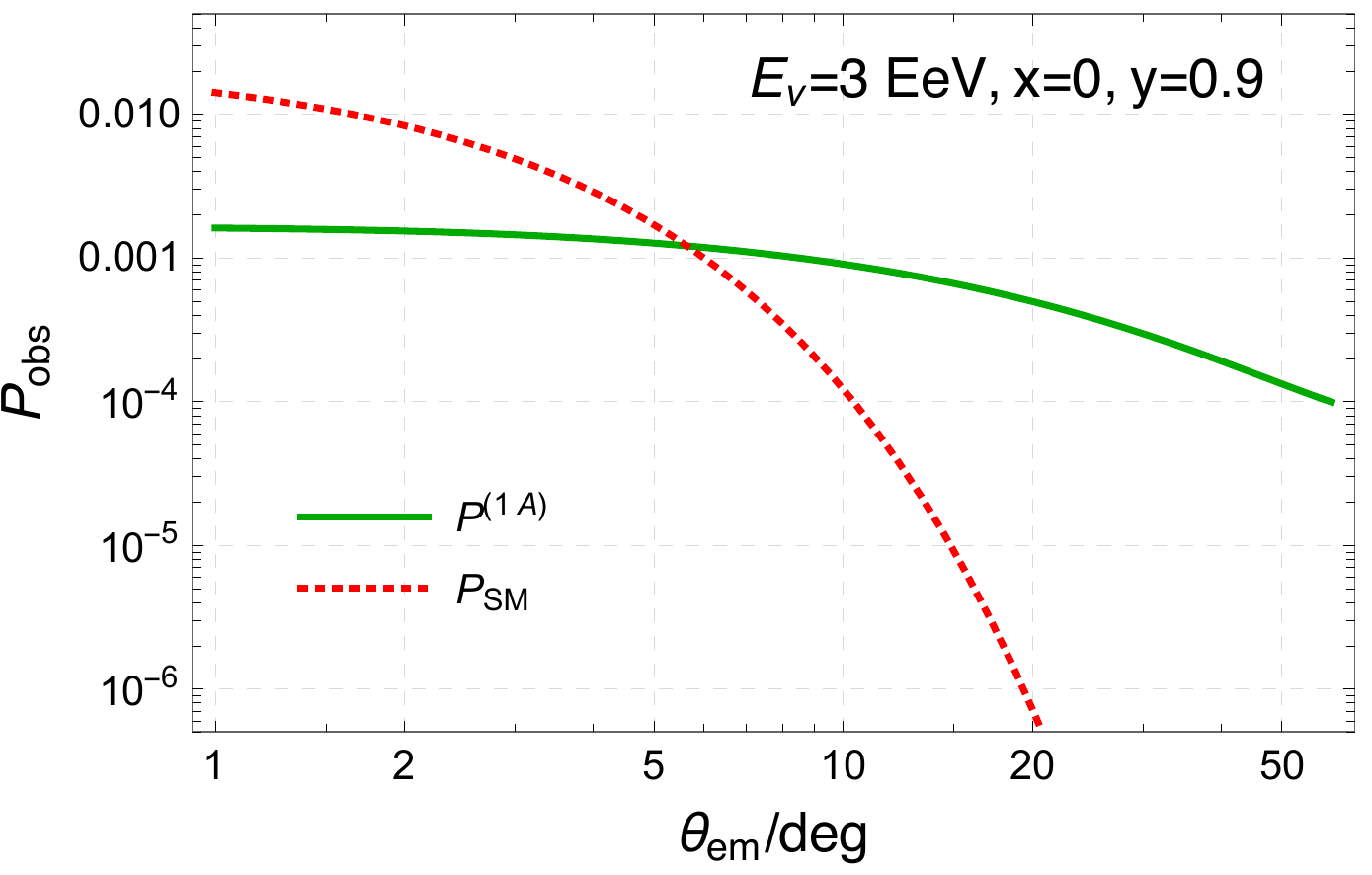}
\hspace{0.01\hsize}
\includegraphics[width=0.485\hsize]{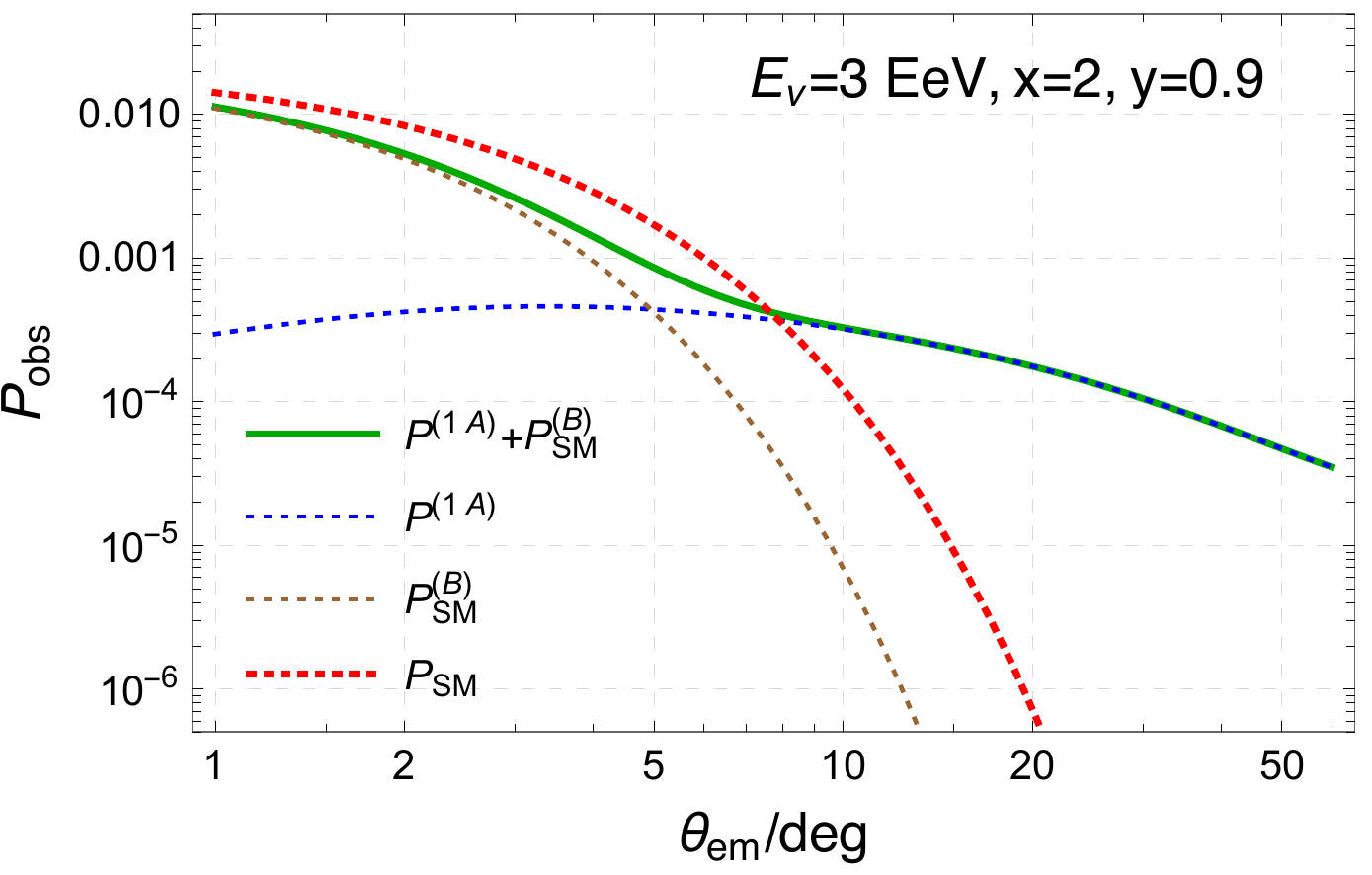}
\caption{Conversion probability (solid green line) in the scenario  \big\langle 1\big\rangle\ $\nu \rightarrow X \rightarrow \tau \rightarrow {\rm hadrons}$, for an incident $3~\mathrm{EeV}$ neutrino to lead to a $\tau$ decaying in the lower 10 km of the atmosphere, as a function of the emergence angle $\theta_{\textrm{em}}$, for a fixed value of the BSM parameters $y=0.9$ and $x=0$ (left) and $x=2$ (right).
For $x \gtrsim 1$, as in the plot on the right, the total conversion
probability is the sum of the process $\nu \rightarrow X \rightarrow
\tau \rightarrow {\rm hadrons}$ (dashed blue line) and the process
$\nu \rightarrow \tau \rightarrow {\rm hadrons}$ (dashed brown line).
For small $x\sim 0$, on the other hand, the process $\nu \rightarrow \tau
\rightarrow {\rm hadrons}$ is too small to show on the scale of
the plot, explaining the absence of $P^{(B)}_{SM}$ in the left-hand
side. For comparison, we show
the conversion probability in the SM (red dashed line).
    }
\label{fig:prob}
\end{figure*}
%

The initial scattering process for BSM physics that converts $\nu$ to $X$ is parametrized by the interaction length
\be
   \lbsm = (\sigma_{\sss BSM} n_\oplus)^{-1}
\ee
where $\sigma_{\sss BSM}$ is the BSM cross section for scattering
of $\nu$ on nucleons (or possibly electrons) and
$n_\oplus$ is the number density of target particles in the Earth.
In case of more than one BSM conversion, such as cascade
decays, the $i$th interaction length is denoted as $\ell_{i}$.
Similarly, the mean free path for the standard model DIS
interaction is denoted by
\be
   \lsm = (\sigma_{\sss SM} n_\oplus)^{-1} \,.
\ee
These interaction lengths generally depend upon $E_\nu$.
\newline

\noindent
{\bf  \big\langle 0\big\rangle\ The SM process $\nu \rightarrow \tau \rightarrow {\rm\bf hadrons}$.} The total conversion probability $P_{\rm SM}$ 
can be expressed as 
\bea
   P_{\rm SM} &=& B_h\int_0^{\ell_{c}} \d z_1 \frac{ \d P_{\nu \rightarrow \tau}  } { \d z_1} 
            \int_{E_{\rm min}}^{E_\nu} \d \bar E_\tau
               \frac{ \d P_{\rm exit} } { \d \bar E_\tau} \nn\\  
          &\times&  \int_{ \ell_{c}}^{ \ell_{c} + \frac{D} { \sin \theta_{\textrm{em}} } }  \d z_3
                  \frac{ \d P_\tau } { d z_3}  \,,
   \label{eq:Psm}
\eea
where $B_h \simeq  0.65$ is the branching ratio for a $\tau$ to decay into hadrons
(we count only the hadronic decays for initiating the kind of air showers
observed by ANITA) and the chord length between the entry point of the UHE
$\nu$ and the exit point of the $\tau$ is  $\ell_c= 2 R_\oplus \sin
\theta_{\rm em}$.  For illustration see the line labeled 0 in fig.\
\ref{fig:path}.
The probability distribution for the incoming neutrino to scatter into a $\tau$
at position $z_1$ is
\bea
   {\d P_{\nu \rightarrow \tau}\over dz_1} &=& 
      \lsm^{-1}\exp\left(-z_1 \lsm^{-1} \right) \,,
\eea
and the probability for $\tau$ to exit with energy $\bar E_\tau$, starting
from initial energy $E_\nu$, is given
by \cite{Feng:2001ue}
\bea
   {\d P_{\rm exit} \over \d \bar E_\tau}&=&  \exp \left(
            \frac{ \ell_\tau^{-1} ( E_{\nu}) - \ell_\tau^{-1} ( \bar {E}_{\tau}) } {  \beta_\tau \rho_\oplus } 
            \right)    
               \nn \\
            &&  \delta\left( \bar {E}_{\tau} - E_{\nu} 
             e^{ - \beta_\tau \rho_\oplus (\ell_c - z_1)} \right)  , 
      \label{eq:Pexit}
\eea
in terms of the $\tau$ decay length $\ell_\tau(E_\tau) = c \tau_\tau
E_\tau/m_\tau$.  
(Here we neglected the fact that the $\tau$ on average receives only 80\% of $E_\nu$ in the interaction \cite{Gandhi:1995tf}.)
The Earth density is taken as a constant for simplicity,
$\rho_{\oplus} = 2.7~{\rm g/cm^3}$, and $\beta_\tau \sim (0.4-0.8)\times
10^{-6}$\,cm$^2$/g in the energy range of interest.  We take
$\beta_\tau = 0.6\times 10^{-6}$\,cm$^2$/g for our estimates.
The probability for the $\tau$ to decay in the atmosphere
at position $z_3$ is
\bea 
   {\d P_\tau\over \d z_3} &=&   \ell_\tau^{-1}\exp\left(-(z_3-
\ell_{c} )\ell_\tau^{-1}\right)  \ . 
\eea
The probability distribution (\ref{eq:Psm}) is integrated over $\bar E_\tau$ between the initial value
$E_\nu$ and a minimum value $E_{\rm min}\sim 0.1\,$EeV that could still be associated with the observed AAEs.

With these analytic formulae, we can reproduce the conversion probability given by the Monte Carlo
simulations in ref.\ \cite{Alvarez-Muniz:2017mpk} with vanishing ice thickness for  $\theta_{\rm em} \lesssim \ang{3}$. 
Unlike the simulations, we neglect $\tau$ regeneration \cite{Halzen:1998be}, which is important for larger $\theta_{\rm em}$
and low $\bar E_\tau$. After exiting the Earth, the $\tau$ does not suffer further energy loss, and it will decay in the atmosphere. 
The resulting probabilities are shown as a function of $\theta_{\rm em}$ in fig.\  
\ref{fig:prob}, in which $P_{\rm SM}$ drops steeply at large $\theta_{\rm em}$.
Even though $\tau$ regeneration increases the probabilities for large
 $\theta_{\rm em}$,  this falls far short of being able to explain the AAEs within the SM~\cite{Fox:2018syq}.
\medskip

\noindent {\bf 
 \big\langle 1\big\rangle\ The BSM process $\bm{\nu \rightarrow X \rightarrow \tau
\rightarrow {\rm\bf hadrons}}$.}
To obtain a higher conversion probability at large $\theta_{\rm em}$, we consider the possibility that the UHE 
$\nu$ can convert to a weakly interacting 
particle $X$ with a larger mean free path than neutrinos, followed by $X$ decaying or rescattering into a $\tau$. 
The conversion probability can be written as 
\begin{eqnarray}
   P^{(1A)} &=& B_h \int_0^{\ell_{c}} \d z_1 \frac{ \d P_{\nu \rightarrow X}  } { \d z_1} 
            \int_{z_1}^{\ell_{c}} \d z_2 \frac{ \d P_{X \rightarrow \tau}  } { \d z_2}
            \nn \\
            &&
            \times\int_{E_{\rm min}}^{E_\nu} \d \bar E_\tau 
               \frac{ \d P_{\rm exit} } { \d \bar E_\tau}   
            \int_{ \ell_{c}}^{ \ell_{c} + \frac{D} { \sin \theta_{\textrm{em}} } }  \d z_3
                  \frac{ \d P_\tau } { d z_3}  
\end{eqnarray}
with superscript $A$ denoting that we neglect the SM contribution to the total conversion
probability here (see eq.\ (\ref{eq:PBsm}) below).  The probability distributions are given by
\begin{eqnarray}
   {dP_{\nu \to  X}\over \d z_1} &=& \lbsm^{-1}\exp\left(-z_1(\lsm^{-1} + \lbsm^{-1})\right)  \ , 
      \\ 
   {dP_{X \to \tau}\over \d z_2} &=& \ell_X^{-1} \exp\left(-(z_2-z_1)\ell_X^{-1}\right)  \ .
\end{eqnarray}
$dP_{\nu \to  X}$ is the differential probability that $\nu$ converts to $X$, determined
by the mean free path for BSM interactions $\lbsm$ (in addition to the damping by the SM contribution
to the $\nu$ conversions).
$\d P_{X \to \tau} $ quantifies the decay or rescattering of $X$ (produced at $z_1$) 
into $\tau$ at $z_2$, with mean free path $\ell_X$.
In the rescattering scenario, there are two possible subcases: the interaction may simply be the inverse of the one 
that produced $X$
(if the original interaction involved the full third generation SU(2)$_L$ doublet $(\nu_\tau,\tau)$
and not just the right-handed component of $\tau$), or it may be a different interaction.
In the first case, one has $\ell_X \cong \lbsm$, so this can be
considered as a special case of the more general situation. The general situation is 
like the second case, where
$\ell_X$ and $\lbsm$ are independent parameters.
$\d P_{\rm exit} / \d E_\tau $ is to account the energy loss of $\tau$ as in eq.~(\ref{eq:Pexit}) with $z_1\to z_2$.

We must also include the SM component, similar to $ \big\langle 0\big\rangle$, 
\bea
   P^{(B)}_{SM}&=& B_h \int_0^{\ell_{c}} \d z_1 \frac{ \d P_{\nu \rightarrow \tau}^\prime  } { \d z_1} 
            \int_{E_{\rm min}}^{E_\nu} \d E_\tau
               \frac{ \d P_{\rm exit} } { \d E_\tau} \nn\\  
           &\times& \int_{ \ell_{c}}^{ \ell_{c} + \frac{D} { \sin \theta_{\textrm{em}} } }  \d z_3
                  \frac{ \d P_\tau } { d z_3}  
   \label{eq:PBsm}
\eea
which is 
however modified by the
new physics because the mean free path for $\nu\to\tau$ is decreased by the BSM
contribution,
\begin{equation}
   {\d P_{\nu \to \tau}^\prime\over \d z_1} = \lsm^{-1}\exp\left(-z_1(\lsm^{-1} + \lbsm^{-1})\right)   \  .
\end{equation}
We omit the number of the process in the superscript of $P^{(B)}_{SM}$, since this same modification will
apply to all three BSM scenarios.

The total probability $P^{(1A)} + P^{(B)}_{SM}$ 
can be expressed in terms of dimensionless variables 
$x$ and $y$,
\be
	x \equiv \lbsm/  \lsm ,\qquad y\equiv \ell_X/ R_\varoplus \,.
\label{xyeq}
\ee
For an incident $3~\mathrm{EeV}$ neutrino to 
generate a $\tau$ decaying in the lower atmosphere, at a fixed value of $\theta_{\rm em}=27^\circ$ 
representative of an AAE,
we find the largest possible probability $\simeq 3.2 \times 10^{-4}$, 
around $x\sim 0$, $y \sim 0.906$. The value of $y$ is close to the chord length  
$\ell_c ( 27^\circ) \simeq  0.906 R_\varoplus$, which makes $X$-decays close to the surface of the Earth likely.  
At $27^\circ$, the BSM probability $P^{(1A)}$ (which is an increasing
function of $1/x$) dominates over $P^{(B)}_{SM}$ by far.

Fig.\ \ref{fig:prob}
illustrates the dependence of $P_{\textrm{\rm obs}}$ on the emergence angle $\theta_{\rm em}$ for the two choices
$(x,y) = (0,0.9)$ and $(2,0.9)$, respectively, and $E_\nu = 3\,$EeV.  
The green (solid) line is the total probability, while the other curves show the individual contributions
$P^{(1,A)}$ and $P^{(B)}_{SM}$, as well as the SM contribution (red dashed curve). 
(Since the latter neglects neutrino regeneration effects, the true SM probability is higher for 
$\theta_{\rm em} \gtrsim 3^\circ$.)
\newline

\noindent{\bf  \big\langle 2\big\rangle\ The BSM process $\bm{\nu \rightarrow X_1 \rightarrow  \cdots \to X_n \to  \tau \rightarrow {\rm\bf
hadrons}}$.} Rather than a single BSM particle $X$ in the conversion,
 it is conceivable to have a cascade of decays
or several consecutive conversions.   
One expects that increasing the number of steps increases the conversion probability, since
the final step can then have a higher likelihood of producing a $\tau$ near the surface  of the Earth.
Going to a very  large number of steps is theoretically unlikely, and part of the initial $E_\nu$ is lost in 
each decay. Therefore we give the numerical results up to
five steps, fixing  $\theta_{\rm em} = 27^\circ$ and $\ell_{c} = 5785~\mathrm{km}$,
by neglecting the energy loss in every conversion, except for the
energy loss of $\tau$, 
taking  the initial $\nu$ energy to be $3~\mathrm{EeV}$.  In table
\ref{tab1}, second column, we show the maximum probabilities as a function of the number of steps
in the cascade.
They are achieved when the conversion rate of $\nu \to X_1$ is high~(small $x$) and all the decay lengths of the $X_i$ are roughly $\ell_i \sim \ell_{c}/ n$. 
The probabilities include the hadronic branching ratio $B_h$.
\newline

\begin{table*}[ht]
\centering
\tabcolsep 8pt
\begin{tabular}{|c||c|c|c||c| c | c| }
\hline
\# steps & max.\ prob.\  \big\langle 2\big\rangle  &  p-value  \big\langle 2\big\rangle &  significance  \big\langle 2\big\rangle  &   max.\ prob.  \big\langle 3\big\rangle  \ &  p-value  \big\langle 3\big\rangle &  significance  \big\langle 3\big\rangle  \\
\hline
1 & $3.2\times 10^{-4}$ &  $3.7 \times 10^{-8}$  & 5.5 $\sigma$ &   $1.4\times 10^{-3}$  &  $5.1 \times 10^{-6}$ & 4.6 $\sigma$  \\
2 & $5\times 10^{-4}$ &   $4.8 \times 10^{-8}$ & 5.5 $\sigma$ & $2.1\times 10^{-3}$  &  $1.5 \times 10^{-6}$  & 4.8 $\sigma$\\
3 & $6 \times 10^{-4}$ & $6.5 \times 10^{-8}$ &  5.4 $\sigma$ &  $2.5\times 10^{-3}$  &  $1.3 \times 10^{-6}$ & 4.8 $\sigma$\\
4 & $7 \times 10^{-4}$ & $8.1 \times 10^{-8}$  &  5.4 $\sigma$ & $3.0\times 10^{-3}$ & $1.4 \times 10^{-6}$  & 4.8 $\sigma$ \\
5 & $7 \times 10^{-4}$ & $9.3 \times 10^{-8}$ &   5.3 $\sigma$ &  $3.3\times 10^{-3}$ &  $1.5 \times 10^{-6}$ & 4.8 $\sigma$ \\
\hline\end{tabular}
\caption{Maximum probability is computed for the AAE with $\theta_{\rm em} = 27^\circ$ ($\ell_{c} = 5785~\mathrm{km}$). 
The $p$-value and significance $\sigma$ take into account the two AAEs.
The values in the process \big\langle 2\big\rangle\ and \big\langle 3\big\rangle\
with different numbers of steps are given.
\label{tab1}}
\end{table*}

\noindent
{\bf\big\langle3\big\rangle\ The BSM process 
$\bm{\nu \rightarrow X_1 \rightarrow  \cdots \to X_n  
\rightarrow {\rm\bf hadrons}}$}.
The AAEs do not necessarily originate from $\tau$ decays. A BSM particle  decaying in the 
atmosphere to hadrons could also give rise to the observed events, for example a dark photon decay
$A' \to h$.  As in case 2, we take $\theta_{\rm em} = 27^\circ$ and $E_\nu = 3~\mathrm{EeV}$, and 
neglect the energy loss of BSM particles while they traverse the Earth. 
The maximum probabilities are shown in column 3 of table \ref{tab1}, where we take the branching ratio
$X_n \to {\rm hadrons}$ as $1$.
These can be approximated by 
an analytic formula, $P_{n} \simeq \sqrt{n/2\pi}\,  
D /(  \ell_{c}\sin \theta_{\rm em})$, which works well
if $\ell_{c}/n \gg D / \sin \theta_{\rm em}$.

In the following, we will refer to the model-dependent 
probability defined in this section generically as 
$P_{\rm obs}(\theta_{\rm em})$, considering it as a function
of the emergence angle.

\subsection{Limits on BSM assuming a diffuse $\nu$-flux} \label{sec:limits}
The predicted number of ANITA events is in general given by 
\cref{eq:expectmu},
where conversion probabilities for different models are given in table \ref{tab1}. 
We use the exposure estimate obtained from the reflective 
events observed by the ANITA-I and III flights \cite{Fox:2018syq},
\be
H_{\textrm{\tiny AN}}^{\textrm{\tiny ref}} \simeq 2.7 \, \textrm{km}^2\, \textrm{yr} \, \textrm{sr}\, \,.
\ee

For the incoming neutrino flux, we take the limits from 
IceCube~\cite{Aartsen:2018vtx} and Auger~\cite{Aab:2015kma}:
the energy times diffuse isotropic flux 
is $\lesssim 6$\,km$^{-2}$\,sr$^{-1}$\,yr$^{-1}$ at $E_\nu=3\,$EeV.  
Assuming that $\nu_\tau$ constitutes 
$1/3$ of the total diffuse flux, due to oscillations, 
this implies
$E_{\nu_\tau}\,\phi(E_{\nu_\tau}) \lesssim 2$\,km$^{-2}$\,sr$^{-1}$\,yr$^{-1}$.
Our conclusions below do not change significantly
on varying the energy in the range of [$0.1\,$EeV, $10\,$EeV].

$H_{\textrm{\tiny AN}}^{\textrm{\tiny ref}}$ does not include the conversion 
probability $P_{\textrm{\rm obs}}$. 
The latter is a function of $\theta_{\textrm{em}}$, so we average over it in order to obtain the estimated total exposure.
As mentioned above we assume that $\d \Omega_d$ does not depend on $\theta_{\textrm{em}}$.
Then the predicted number of events from a diffuse flux in
the region visible to ANITA can be estimated as
\begin{equation}
   \mu \simeq \frac{ \int \d A P_{\rm obs} ( \theta_{\textrm{em}} ) 
\sin \theta_{\textrm{em}} }  { \int \d A   
\sin \theta_{\textrm{em}} }  \, 
H_{\textrm{\tiny AN}}^{\textrm{\tiny ref}}  \, E_\nu\,\phi(E_\nu) \ .
\label{mueq}
\end{equation}

To check whether the AAEs can be compatible with the BSM scenarios discussed above we perform an extended likelihood analysis. 
The extended likelihood takes into account the total number of observed AAEs $n$ as well as the measured emergence
angles $\theta_{\textrm{em}}^i$ of these events, and is defined as 
\begin{equation}
   L(n,\theta_{\rm em}) = \frac{e^{-\mu}\mu^n}  { n !} \prod_{i=1}^n 
   f( \theta_{\textrm{em}}^i)   \  , 
\end{equation}
where
\begin{equation}
   f ( \theta_{\textrm{em}} ) = \frac{ P_{\rm obs}(\theta_{\textrm{em}})
               \sin\theta_{\textrm{em}}
             \sin \theta_{E } \,\d \theta_E 
 /\d \theta_{\textrm{em}} 
            }   {     \int  \d \theta_E \sin \theta_{E } P_{\rm obs} ( \theta_{\textrm{em}} )  \sin \theta_{\textrm{em}}      } \ ,
\end{equation}
is the probability distribution function  to observe the emergence 
angle $\theta_{\textrm{em}}$, and $L(0,\theta_{\rm em})\equiv 1$.
From $L$ one can compute the $p$-value
\be
	p = \sum_{n=0}^\infty \int 
      L(n,\theta_{\rm em}) \,
         \prod_{i=1}^{n} d\theta^i_{\rm em}
\label{pval}
\ee
for ANITA to observe at least two events with the respective emergence 
angles $27^\circ$, $35^\circ$.  The angular integration region is 
defined to be such that 
\be
P(n;\mu)\prod_{i=1}^n f(\theta_{\rm em}^i) < 
P(2;\mu)f(27^\circ)
f(35^\circ)
\ee
in terms of the Poisson distribution $P(k,\mu) = \mu^k e^{-\mu}/k!$.
In practice, the important contributions to (\ref{pval})
come from $n\ge 2$, and for $n\ge 3$ one can neglect the angular
dependence and take $L(n) = P(n;\mu)$.
For BSM model  \big\langle 1\big\rangle, we find that the parameters $x$ and $y$ that
 maximize the 
extended likelihood
are $x \sim 0 $, $y \sim 1.13$. 
As expected $y$ is slightly different from the one given in the previous section since we are considering two events with emergence angles $27^\circ$ and $35^\circ$, respectively.   
The corresponding $p$-value is $3.7\times 10^{-8}$, which
implies that the BSM topology \big\langle 1\big\rangle\ is inconsistent with the data at 
the level of $5.5\, \sigma$. As shown in \cref{tab1}, for the model \big\langle 2\big\rangle, the discrepancies
are roughly the same, $\sim 5.5\, \sigma$; for the model \big\langle 3\big\rangle, the discrepancy can be
reduced to the level of $4.6\, \sigma$. 
It turns out that increasing the number of steps does not 
correspond to an decrease of the $p$-value, which depends upon
the total acceptance and not just on $P_{\rm obs}$.

We conclude that if a diffuse neutrino flux is assumed to be the
source,  BSM physics is insufficient to explain the ANITA anomalous
events. The minimum incompatibility, tension at the level of
4.6\,$\sigma$,  is obtained for a cascade of several BSM particles 
$X_i$ in which the last one directly decays to hadrons
with a branching fraction close to unity.

\subsection{Limits on BSM from point sources}

The IceCube diffuse flux upper limit for $\nu_\mu$ (that we
presume to also hold for $\nu_\tau$), $E_\nu\,\phi(E_\nu) \lesssim 2$\,km$^{-2}$\,sr$^{-1}$\,yr$^{-1}$, 
rules out both SM and BSM explanations of the AAEs. 
A loophole could in principle be bright neutrino point sources.
If there exists a large number of
point sources with an isotropic distribution, 
their total flux cannot exceed the diffuse 
flux limit. Thus for an isotropic distribution of many point sources, one reaches the same conclusion as for the diffuse flux. 
On the other hand, when the number of point sources $N_{\rm ps}$ is low, 
the single source flux limit is bounded by
total diffuse flux divided by $N_{\rm ps}$; see fig.~$8$ of 
\cite{Aartsen:2018ywr}.
The optimal situation for explaining the AAEs is then to have
only two points sources, both in the Northern hemisphere, that give
rise to the two AAEs.

For a single point source of muon neutrinos, assuming a spectrum of the form
$\phi_{\rm ps}\sim E^{-2}$, the sensitivity of IceCube is 
$\phi_{\rm ps}\, E_\nu = 0.13 \, \textrm{km}^{-2}\,\textrm{yr}^{-1}$ 
at $E_\nu = 1\,\rm{EeV}$
\cite{Aartsen:2018ywr}.  The IceCube sensitivity to $\nu_\tau$ is
lower because of poorer angular resolution, but we will assume that
the two fluxes are equalized by oscillations.
Since the angle is fixed for point sources, the unit of flux does not
include $\textrm{sr}^{-1}$, and this changes the estimated 
number of observed events from eq.\ (\ref{mueq}) to the form
\begin{eqnarray}
   \mu_{\rm ps} &\simeq& \int \d A_{ps} P_{\rm obs}(
\theta_{\textrm{em}} ) \d t  \, \times \phi_{ps}\, \d E_\nu
   \\
      &\simeq & \frac{ P_{\rm obs}( \theta_{\textrm{em}} )    A_{\rm ps} } 
 {  \int \d A_{\rm ps} \sin \theta_{\textrm{em}}  \d \Omega_d } \times 
      H_{\textrm{\tiny AN}}^{\textrm{\tiny ref}} \times  \phi_{\rm
ps}\, E_\nu
   \nonumber
   \\ 
      &\simeq & \frac{ P_{\rm obs}( \theta_{\textrm{em}} )  \left( h / \sin \theta_{\textrm{em}} \right)^2  }  
{  \int \d A_{\rm ps} \sin \theta_{\textrm{em}} } 
            \times H_{\textrm{\tiny AN}}^{\textrm{\tiny ref}} \times 
\phi_{\rm ps}\, E_\nu \ ,    \nonumber
\end{eqnarray}
where $\d A_{\rm ps}  =  ( h / \sin \theta_{\textrm{em}} )^2 \d 
\Omega_d$
is the differential area over which ANITA observes an air shower 
originating from a point source at an angle $\theta_{\textrm{em}}$
and $h \cong 35\,$km is its altitude.
Taking $ \int \d A_{\rm ps} \sin \theta_{\textrm{em}}   \simeq 9\times
10^4 ~\mathrm{km}^2$, we can estimate in the BSM model  \big\langle 1\big\rangle, the
maximum $\mu_{ps} = 1 \times 10^{-5}$ for two point sources in the
northern hemisphere, excluding the model at the $6.5\,\sigma$
confidence level. The reason for this stronger exclusion is
the assumption of the spectrum 
$\phi(E_\nu)\sim E_\nu^{-2}$ \cite{Aartsen:2018ywr}. It leads to fewer
events at $E_\nu \sim 1\,$EeV from anisotropic point 
sources than does the diffuse flux.

This suggests another possible loophole (apparently taken
in ref.\ \cite{Collins:2018jpg}), namely to have a non-monotonic
spectrum with a peak at energies $\sim 1$\,EeV.  Ref.\ \cite{Aartsen:2018ywr}
computes the sensitivity and discovery potential of IceCube to
such sources, as a function of $E_\nu$ and declination $\delta$ (see
fig.\ 4 in that reference).
For $\delta = 30^\circ$, similar to the angles of the AAEs, the 
differential flux limits are 
$\phi_{ps}\, E_\nu \sim 6\times 10^{3}\,{\rm km}^{-2}\,{\rm yr}^{-1}$ 
at $E_\nu = 0.5\,{\rm EeV}$ and 
$\phi_{ps}\, E_\nu \sim 1.6\times 10^{5}\, {\rm km}^{-2}\,{\rm yr}^{-1}$
at $E_\nu = 1\,{\rm EeV}$. Taking the former flux we find that  
BSM model  \big\langle 1\big\rangle\ is in tension with the ANITA data at only $1.6\,\sigma$,
while with the latter, all the models including the SM are 
consistent with the data.

However at declination $\delta = 30^\circ$ and $E_\nu\sim 1$\,EeV,
the Auger experiment is more sensitive than IceCube, since events from
this direction are down-going or Earth-skimming \cite{2012ApJ...755L...4P}.  A
limit  of $E_\nu^2\,\phi_{ps} < 140\,$EeV\,km$^{-2}$\,yr$^{-1}$ is
reported for an $E_\nu^{-2}$ flux from a point source.  For a flux
concentrated at $E_\nu\sim 1\,$EeV, the limit is presumably weaker,
but probably not greatly so since Auger's sensitivity to differential
spectra is peaked around 1\,EeV. It is reasonable to assume the differential limit at $\sim 1\,$EeV is 
a factor of $5$ weaker than the $E_\nu^2$ flux. Taking the differential limit of 
$E_\nu \phi_\nu < 700$\,km$^{-2}$\,s$^{-1}$ at $E_\nu\sim 1\,$EeV,
this would be in $3.1\,\sigma$ tension for
being able to explain the AAEs.  We conclude that nonpower-law flux
point sources at high declination, while perhaps not ruled out, 
can only be marginally consistent with the AAEs.

Finally, intense transient sources lasting $\ll$ 24 hours, 
coming from the northern hemisphere and peaked in the EeV range, might
have been missed by both Auger and IceCube, so that we are unable
exclude this as a possible explanation, either within the SM or
coming from new physics.  A study of this possibility is beyond the
scope of the present work.

\section{ANITA events from decaying Dark Matter}
\label{sec:decayDM}

Since very large fluxes are required to overcome the small
transmission probability through the Earth, one is motivated to look
for an exotic source of BSM particles that could produce AAEs, whose
flux could be much larger than the experimental limits on the 
neutrino
diffuse or pointlike fluxes.  Decaying dark matter has been suggested.  In one
scenario, the signal is dominated by decays of DM that has been
trapped in the Earth \cite{Anchordoqui:2018ucj}, while in another it
is DM decaying in the galactic halo that dominates
\cite{Heurtier:2019git}.  We will  argue that both of these are in 
varying degrees of tension with complementary constraints. However it is possible that decaying
DM gives rise only to highly boosted BSM particles, that interact in the
Earth to produce the AAEs. We point out in the following Section that this third scenario,
heretofore unconsidered, can be viable.

\subsection{DM decays in the Earth}

DM particles $\chi$ can accumulate inside the Earth if they have a
sufficiently large cross section $\sigma_{\chi N}$ for scattering on nucleons.  
The current upper limit  from XENON1T,
extrapolated to DM mass $m_\chi = 1$\,EeV, is $\sigma_{\chi N} <
10^{-39}\,$cm$^2$ \cite{Aprile:2018dbl}, 
giving a mean free path in the Earth of 
$\lambda > 0.3\times 10^{10}$\,km, which is $5\times 10^5$ times
greater than the radius of the Earth.  The collection efficiency
is expected to be $\epsilon = 
\pi R_\oplus/(2\lambda)\lesssim 3\times 10^{-6}$ (accounting for the 
average distance through the Earth traversed by a DM particle).  
The total mass of DM which the
Earth can collect is
\be
M_\chi = \epsilon \, \rho_{\textrm{\tiny DM}}^{\textrm{local}} \, 
v_\textrm{rel} \,  \pi R_\varoplus^2 \, t_\varoplus  \lesssim 
6 \times 10^{12}{\rm\, g} \,,
\ee
where  $\rho_{\textrm{\tiny DM}}^{\textrm{local}} \simeq 0.3 \,
\textrm{GeV}/\textrm{cm}^3$ is the local DM density, $v_\textrm{rel}
\simeq 220 \, \textrm{km/s}$ is the DM velocity and 
$t_\varoplus \simeq 4.5$\,Gyr is the age of the Earth.

On the other hand, in order to explain the $N_\textrm{\tiny A}$=2
anomalous events within $t_\textrm{\tiny A}=$ 34 days, in a solid angle
$\Omega_\textrm{\tiny A} \simeq 4 \textrm{\,km}^2/(4 \pi R_\varoplus^2)$,
and with DM 
lifetime $\tau_\chi$, 
one needs a total DM mass accumulation of
\be
M_\chi = \frac{m_\chi \tau_\chi N_\textrm{\tiny A}}{t_\textrm{\tiny A}
\Omega_\textrm{\tiny A} P_\textrm{\tiny A}}
 \simeq  2\times 10^{17} {\rm\, g} \left(
\frac{\tau_\chi}{2\times 10^{27}\,\rm{s}} \right)
\left( \frac{10^{-3}}{P_\textrm{\tiny A}}\right) \,,
\ee
where $P_\textrm{\tiny A}$ is the probability for a DM decay
product to propagate through the Earth and produce a 
$\tau$ that initiates a hadronic shower in the lower atmosphere.   
We take an optimistic
value of $P_\textrm{\tiny A}\sim 10^{-3}$ based on our previous 
estimates.  The lifetime is constrained by  IceCube
searches for neutrinos from decaying DM \cite{Aartsen:2018mxl} to be
$\tau_\chi\gtrsim 2\times 10^{27}$\,s.  This is the weakest limit
(corresponding to $\chi\to\nu\nu$)
which in general depends upon the final state particles in the decay. 
  The required mass is
five orders of magnitude greater than
the amount of DM that could collect inside the Earth, so even with
an unrealistically large value for the conversion probability
$P_\textrm{\tiny A}\sim 1$, the scenario is still ruled out.

One might ask whether this negative conclusion could be evaded by
considering strongly interacting dark matter, that is stopped in the
Earth before reaching underground detectors.  However the larger cross
sections needed have been excluded by other experiments and
astrophysical constraints, as summarized
in ref.\ \cite{Mack:2007xj}.  That reference further closes any remaining
loophole by showing that too much heat would be deposited in the Earth
relative to the measured heat flow, if the direct detection bound is
significantly violated; hence our upper bound on
$\epsilon$ is robust.

\subsection{DM decays in the galactic halo}

Even without collecting in the Earth, decays in the DM halo of our 
galaxy can provide a flux of sterile BSM particles that is far bigger 
than the experimental limit on the diffuse neutrino flux.  This  could
potentially overcome the small survival probability for traversing the
Earth, if these BSM particles can convert to $\nu_\tau$, $\tau$ or
some other particle that produces a hadronic shower in the
atmosphere.  This is the strategy of ref.\ \cite{Heurtier:2019git}, in
which a bosonic DM particle $\chi$ is presumed to decay into sterile
right-handed neutrinos $\nu_R$, that have a small mixing angle $\theta
\sim 0.01$ with active neutrinos. Through this mixing, $\nu_R$ can
interact with nucleons in the  Earth to produce $\nu_\tau$, leading to
the observed AAEs.

However because of the mixing, there are also direct decays
$\chi\to\nu_R\nu_\tau$, with a branching ratio of $\theta^2$,
rendering the decays themselves detectable by IceCube, regardless of
$\nu_R$ interactions in the Earth.  Ref.\ \cite{Heurtier:2019git}
explains the observed AAEs with $m_\chi = 20\,$EeV and $\tau_\chi =
\theta^2\times 10^{27}$\,s.  The effective lifetime for the 
subdominant decays $\chi\to\nu_R\nu_\tau$ is therefore simply
$10^{27}$\,s, independent of $\theta$.  This is in tension with IceCube
limits on decaying DM in the $\chi\to\nu\nu$ channel 
\cite{Aartsen:2018mxl}.  Although the latter reference stops slightly
short of $m_\chi = 1\,$EeV, the limits have been extended to higher
masses \cite{Kachelriess:2018rty}, giving $\tau_\chi > 2\times 10^{28}$\,s, at $m_\chi =
20\,$EeV.
As a consequence, the expected number of AAEs is at least an order of magnitude too low.
Ref.~\cite{Heurtier:2019git} also identifies a region of parameter space around $m_\chi = 5\times
10^4\,$EeV, $\tau_\chi = \theta^2\times 10^{25}\,$s, corresponding to
an effective lifetime of $10^{25}\,$s for the $\chi\to\nu_R\nu_\tau$
channel.  This too is disfavored, since the limits on the lifetime from~\cite{Kachelriess:2018rty} are two orders of magnitude stronger than what one would need to obtain an order one number of AAEs.

\section{A viable model: DM decays to DM}

The previous example suggests that one might achieve a viable
explanation if DM decays solely into a BSM particle whose interactions
in the Earth can produce an exiting $\tau$, with sufficiently high
probability.  Here we present a phenomenological model with a heavy
metastable DM particle $\Psi$ decaying to a highly boosted, light DM
particle $\chi$, that can interact in the Earth to produce $\tau$
leptons.  

The heavy component $\Psi$ must have a mass $m_\Psi\sim 1$\,EeV to
produce the AAEs.  This is above the mass limit for thermal production
of DM, but there exist several mechanisms for producing super-heavy
DM, for example by gravitational particle production 
\cite{Kuzmin:1998uv,Chung:1998zb,Kannike:2016jfs} or
reheating/preheating \cite{Kuzmin:1997jua,Chung:1998ua,Greene:1997ge,Chung:1998rq}
at the end of inflation, freeze-in \cite{Kolb:2017jvz}, dilution by
entropy production \cite{Bramante:2017obj}, or bubble collisions
during a phase transition \cite{Chung:1998ua}.  We suggest a scenario
where $\Psi$ may be a subdominant component of the total dark matter,
while the lighter $\chi$ particle constitutes most of the DM and gets
its relic density from thermal freezeout.

\subsection{Particle physics model}

For a specific model, we take both $\Psi$ and $\chi$ to be fermionic, 
with $\chi$ coupling to lepton doublets,
in particular $L_\tau$, and an inert Higgs doublet $\phi$.  The
interaction Lagrangian is
\be
	{1\over \Lambda^2}\,(\bar\chi \chi^c)\,
	(\bar\chi\Psi) + y_\tau\bar L_\tau \phi \chi
	+ {\rm h.c.}
\label{nlag}
\ee
There is a $Z_4$ charge carried by the new particles that guarantees
the stability of $\chi$.  The scale of the dimension-6 operator is
related to the lifetime of $\Psi$ by
\be
	\Lambda = 2\times 10^{21}\,{\rm GeV} \left(m_\Psi\over 3\,{\rm EeV}
	\right)^{5/4}\left( \tau_\Psi\over 5\times 10^{17}\,{\rm s}
	\right)^{1/4} \,.
\ee
(If desired, a generalization of the model that brings $\Lambda$ below the Planck
scale is to introduce a $Z_{2n}$ discrete symmetry under which
\bea
	\Psi &\to& e^{i(2n-1)\pi/(2n)}\Psi,\nn\\
	\chi &\to& e^{i\pi/(2n)}\chi,\nn\\
	\phi &\to& e^{-i\pi/(2n)}\phi,
\eea
so that $\Psi\to (2n-1)\chi$ through
an operator of dimension $3n$, $\bar\psi\chi (\bar\chi^c\chi)^{n-1}$,
with $n\ge 3$.)

%
\begin{figure}[t]
\includegraphics[width=0.5\hsize]{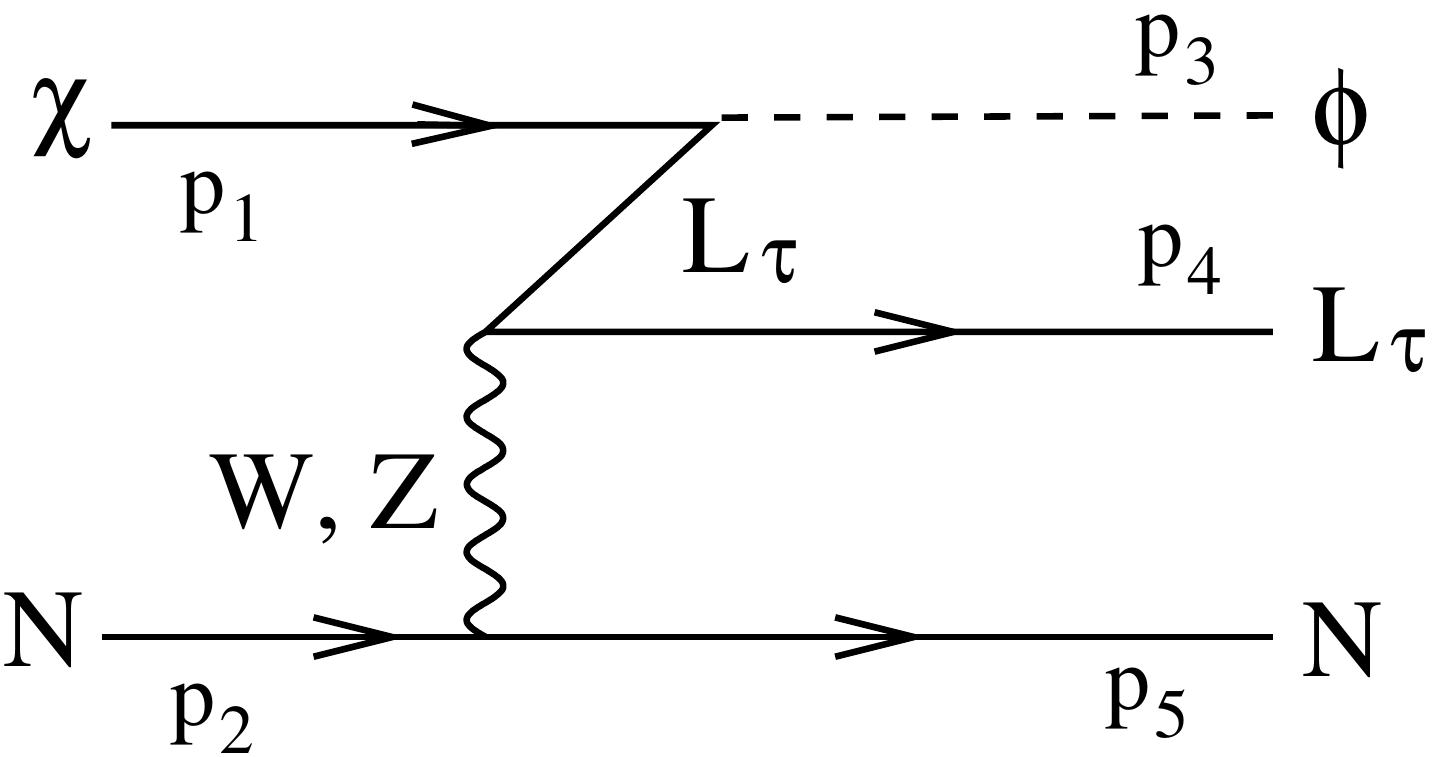}
\caption{Interaction of boosted dark matter $\chi$ with nucleons $N$.} 
\label{fig:dm}
\end{figure}
%

For superheavy DM with $m_\Psi\sim 3\,$EeV, the decay product 
$\chi$s are
highly energetic, and can interact with nucleons in the Earth by the
diagram shown in fig.\ \ref{fig:dm}.  A novel feature is that the
internal $L_\tau$ can go on shell, so the cross section is
logarithmically sensitive to the width of $\nu_\tau$, $\Gamma_\nu$.  
Since the
decay rate of $\nu_\tau$ is negligible, this width is dominated by the
interaction rate of $\nu_\tau$ in the Earth,
\be
	\Gamma_\nu = n_N\sigma_{\nu N} \cong 5\times 10^{-22}\,{\rm
GeV}\left(E_\nu\over {\rm EeV}\right)^{0.3}
\ee
where $n_N\cong 3.3\times 10^{24}/$cm$^3$ is the density of nucleons
and the energy dependence of $\sigma_{\nu N}$, valid around $E_\nu\sim $EeV, is taken
from ref.\ \cite{Alvarez-Muniz:2017mpk}. 

At high energies, the scattering is dominated by low-virtuality $W$
or $Z$ exchange.  We find that the charged-current cross section is
(see appendix \ref{appA} for details)
\be
	\sigma \cong {G_F^2\,m_W^2\, y_\tau^2 \over 64\,\pi^3}
	\ln\left(\sqrt{\hat s}\over \Gamma_\nu\right)
\label{sigma-p}
\ee
where $\hat s = x s$ is the parton-level invariant, with
momentum fraction $x$.  The function $\ln(\sqrt{\hat
s}/\Gamma_\nu)$ denotes the leading behavior at large values of the
argument, but we use the more exact expression (\ref{lastint}) for the
following estimates.  

\begin{figure}[t]
\includegraphics[width=0.95\hsize]{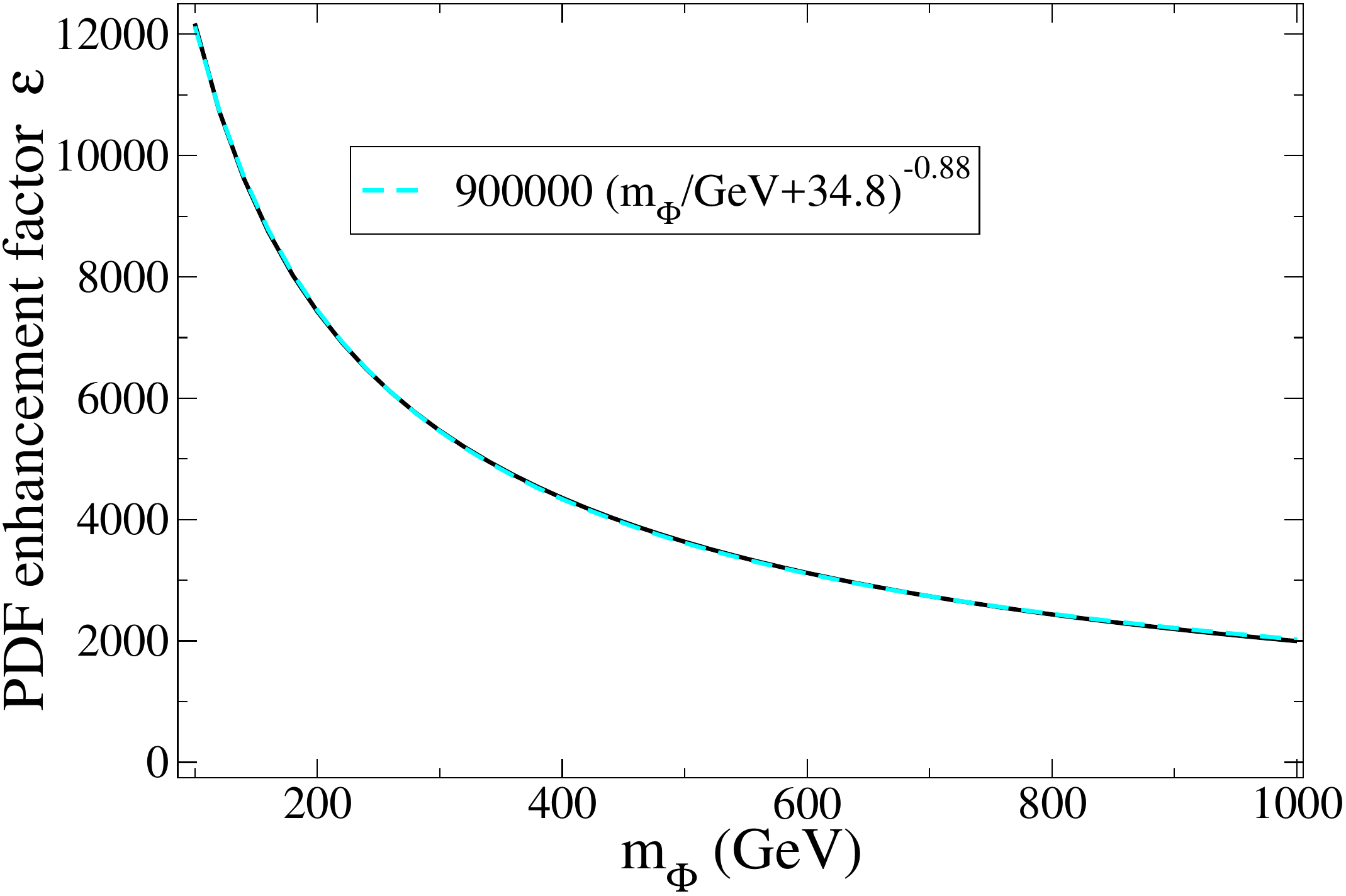}
\caption{Enhancement factor of the $\chi N$ cross section from
integration over parton distribution functions.  The analytic
fit (dashed curve) coincides with the numerical result.} 
\label{fig:E}
\end{figure}

Averaging (\ref{sigma-p}) over the parton distribution
functions of the nucleons gives a result that depends upon the 
inert Higgs doublet mass $m_\phi$, because the cross section is 
dominated by small $x$, whose minimum value is $x_{\rm min} =
m_\phi^2/s$.  The enhancement factor ${\cal E}$ takes the place
of $\ln(\sqrt{\hat
s}/\Gamma_\nu)$,
\be
	\langle\sigma\rangle = 
{\cal E}\,{G_F^2\,m_W^2\, y_\tau^2 \over 64\,\pi^3}
\label{avgsigma}
\ee
and ${\cal E}$ varies from $12,000$ for $m_\phi=100\,{\rm GeV}$ to
$3,000$ for $m_\phi=1\,{\rm TeV}$, as shown in fig.\ \ref{fig:E}.
We find that it is accurately described by the analytic fit
\be
{\cal E} \cong 9\times 10^5\,(m_\phi/{\rm GeV} + 35)^{-0.88}\, .
\label{efit}
\ee

%
%

\subsection{ANITA anomalous events}

Our model can be considered to be of the type  \big\langle 1\big\rangle\ in our
classification of section \ref{sec:conversion}, in the limit where
$x \to 0$ (see eq.\ (\ref{xyeq})), which indicates that the BSM
particle $X = \Psi$ is already present upon entry of the Earth. The
scattering process actually produces two sources of $\tau$: first from
the primary vertex giving the doublet $L_\tau$ shown in fig.\ \ref{fig:dm},
and second through the fast decays of the produced $\phi\to \chi
L_\tau$.  In section \ref{sec:conversion} we showed that the
probability for these $\tau$s to be observed is maximized when  $y =
\ell_x/R_\oplus\cong 1$.  In the remainder we will assume this
condition is satisfied, to partly  constrain the parameter space of
the model.

\begin{figure}[t]
\includegraphics[width=0.95\hsize]{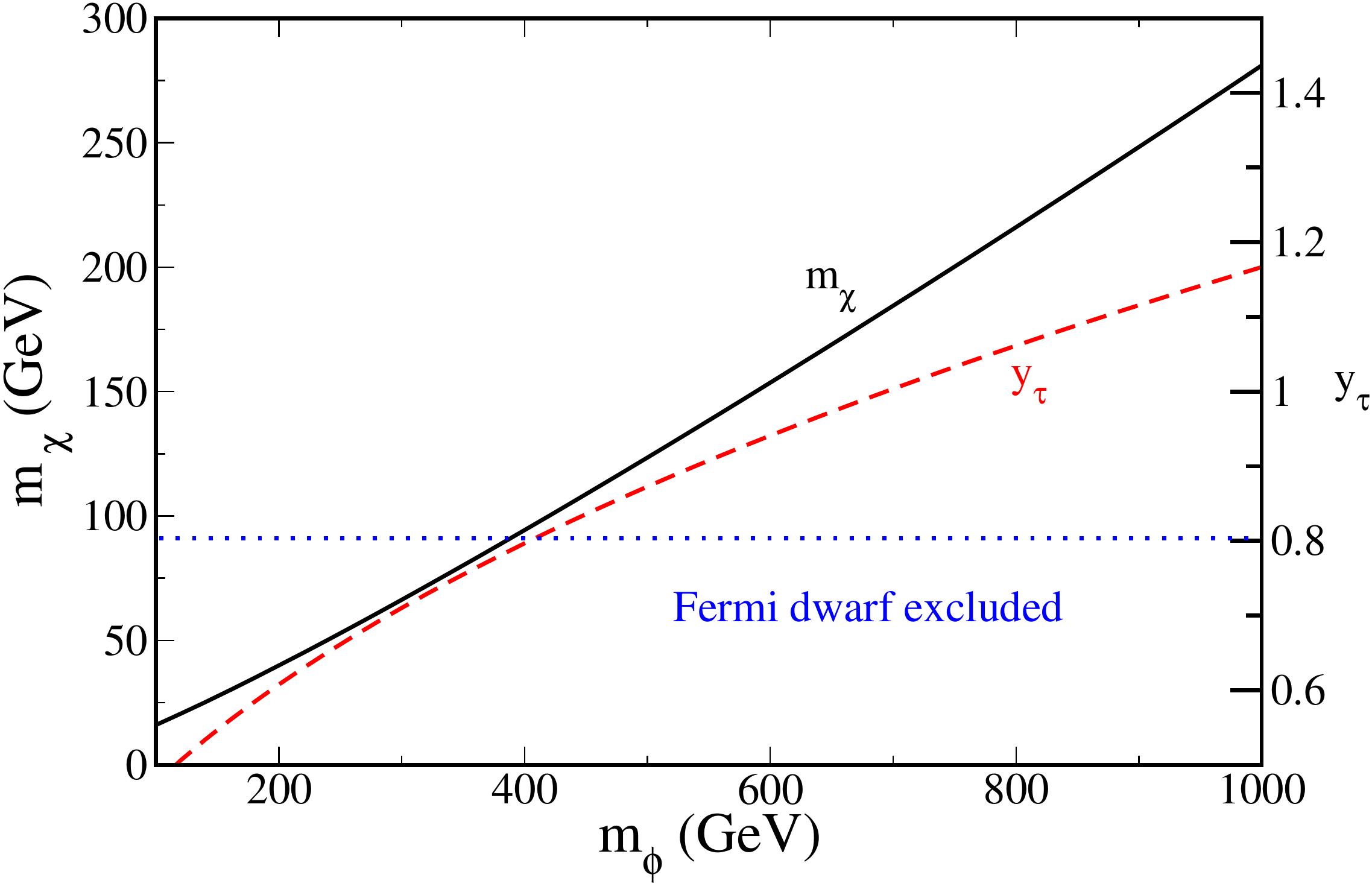}
\caption{Dashed curve: new Yukawa coupling $y_\tau$ versus
$m_\phi$ needed to fulfill $y = \ell_\chi/R_\oplus = 1$.  Solid
curve: value of $m_\chi$ versus $m_\phi$ that gives the correct
relic density for DM $\chi$.
} 
\label{fig:mchimphi}
\end{figure}

The cross section  (\ref{avgsigma}) leads to a scattering length in the
Earth of
\be
	\ell_\chi = {y_\tau^{-2}}\left(3000\over{\cal
E}\right) \times 5700\,{\rm km}
\ee
taking the density of nucleons to be $n_N = 3.3\times 10^{24}$/cm$^3$.
Imposing $y=1$ fixes $y_\tau$ as a function of $m_\phi$ via the
relation (\ref{efit}).
We achieve lengths of order the Earth radius with reasonable
values of $y_\tau\lesssim 1.2$, as shown in fig.\ \ref{fig:mchimphi}.

The procedure of section
\ref{sec:conversion}, with $x=0$ and $y=1$ in process  \big\langle 1\big\rangle\
gives a probability of $P\sim 5\times
10^{-4}$ at large $\theta_{\rm em}$, like for the AAEs.
Then using the likelihood method of section \ref{sec:limits}, we
find that the best-fit flux of $\chi$ particles is $\Phi_\chi =
2230\,$km$^{-2}$\,yr$^{-1}$\,sr$^{-1}$.  For a rough estimate, we can
compute the flux coming from DM decays by assuming that there is a
constant mean density of $\chi$,  $n_\chi = 3 n_\Psi 
(\tau_u/\tau_\Psi)$
from the decays, where $n_\Psi = \rho_\Psi/m_\Psi = f_\Psi\Omega_{\rm cdm}
\rho_{\rm crit}/m_\Psi$ and  $\tau_u$ is the age of the universe. 
Here $f_\Psi\ll 1$ is the abundance relative to that if $\Psi$
constituted all of the DM.  
 
The isotropic flux is then
\be
	\Phi_\chi \sim {n_\chi c\over 4\pi} = 8\times
10^{11}\,f_\Psi\left(\tau_u\over\tau_\Psi\right)\, \left(
  \frac{ 3\, {\rm EeV}}{ m_\Psi}\right) \,   {\rm km}^{-2}\,{\rm yr}^{-1}\,
	{\rm sr}^{-1}
\ee
which determines the $\Psi$ lifetime when equated with the best-fit
flux:
\be
	\tau_\Psi \sim f_\Psi \times 10^{26}\,{\rm s} \,.
\ee
A more quantitative estimate using the $J$-factor for decays in the
galactic halo gives a similar estimate.
One needs
a very high scale $\Lambda \sim f_\Psi^{1/4}\times 10^{24}$\,GeV in the
dimension-6 operator, that could be lowered by taking a larger value
of $n$ in the generalized version of the model.
  We do not concern ourselves here with trying to
build a UV complete model, but instead emphasize that the general
framework may be promising for understanding the AAEs via new physics.

%
%

\subsection{Relic density and other constraints}

It is interesting  that the same interaction that induces
the scattering of $\chi$ in the Earth can also determine its thermal
abundance through the annihilations $\chi\bar\chi\to L_\tau
\bar L_\tau$.  The cross section at threshold is
\be
	\sigma_{\chi\bar\chi}v_{\rm rel} = {y_\tau^4 m_\chi^2\over 
	16\pi\,(m_\phi^2 + m_\chi^2)^2} \,.
\ee
As a rough estimate, we equate this to the nominal level
$(\sigma v)_0 \cong 2\times 10^{-26} \mathrm{cm}^3/\mathrm{s}$ that leads to the observed
relic density.  In conjunction with the determined value of $y_\tau$
for fitting the AAEs, this imposes a relation between $m_\chi$ and
$m_\phi$ that is plotted as the solid curve in fig.\ 
\ref{fig:mchimphi}.  It is approximately fit by $m_\chi \cong
0.3\,m_\phi - 22\,$GeV, consistent with the requirement that $m_\phi >
m_\chi$ so that $\phi\to\chi L_\tau$ decays occur. 

The inert doublet $\phi$ could be pair-produced though electroweak
interactions at the LHC, leading to $\tau$ pairs with missing energy.
This is the same signature as for stau pairs in the MSSM that decay
to $\tau$s and neutralinos, $\tilde\tau\to \tilde\chi^0 \tau$.  ATLAS 
has searched for this
signal in the 8\,TeV data \cite{Aad:2015eda} and more recently CMS has searched
in the 13\,TeV run \cite{Sirunyan:2018vig}.  So far the sensitivity does not reach
the expected production cross section, though in a future run at the
HL-LHC it is projected that $m_\phi$ (in the guise of $\tilde\tau$) will be
excluded below 650\,GeV \cite{CMS:2018imu}.

The annihilation channel $\chi\bar\chi\to \tau^+\tau^-$ is constrained
by Fermi observations of gamma rays from dwarf spheroidal galaxies
\cite{Ackermann:2015zua}.  These exclude $m_\chi \lesssim 80$\,GeV, as indicated
on fig.\ \ref{fig:mchimphi}.

\section{Conclusions}

The origin of the two anomalous events observed by the ANITA balloon
experiment remains unexplained. In this paper we explored if and how
these AAEs might be explained by new physics, within the assumption that BSM states transverse the Earth
and converts to $\tau$ or hadrons in the atmosphere.  
Our conclusions are:

a) A diffuse flux of $\nu_{\tau}$ cannot explain the AAEs, and it is disfavored by $\sim 5 \sigma$
by considering single conversion and cascade decay models. For point sources, the tension is reduced 
to $\sim 3 \sigma$, since the flux limit is weaker.

b) DM decaying inside the Earth cannot account for the AAEs because not enough DM could have
been accumulated inside the Earth during its history.

c) A large flux of ``sterile'' BSM particles whose interactions in the
Earth can produce a $\tau$ near the Antarctic surface could in
principle explain the AAEs. Such a flux of sterile particles could
originate from the decay of DM in the galactic halo. One must ensure
however that the DM decay does not also induce a flux of non-sterile
particles, such as active neutrinos, in excess of the limits from 
IceCube. As an
illustration of a viable model, we discuss the case of a metastable
EeV-scale dark matter particle that decays exclusively to a 
lighter dark matter particle
$\chi$. The latter can in turn scatter within the Earth to produce a
$\tau$ air shower, consistently with all constraints. The relic
density of the stable DM can be explained by the same interaction
that induces the air showers, which can also lead to events at the
LHC resembling supersymmetric $\tilde\tau$ production and decay.

\begin{acknowledgments}
The authors would like to thank Markus Ahlers, Kfir Blum, Jack Collins, Derek Fox, Peter Gorham, Joachim Kopp,
Kohta Murase, Matteo Puel, Sarah Shandera, Ian Shoemaker
and Donglian Xu for useful discussions. 
CG and WX are supported by the European Research Council grant NEO-NAT.
JC is supported by the Natural Sciences and Engineering Research
Council of Canada.
\end{acknowledgments}

\begin{appendix}

\section{DM-nucleon cross section}
\label{appA}
Here we give some details of the computation of the 
charged-current scattering process
shown in fig.\ \ref{fig:dm}.  The spin-averaged squared matrix
element takes the simple form
\be
	\langle|{\cal M}|^2\rangle = {y_\tau^2\,g^4\over 2} 
	{(p_1\cdot p_3)(p_2\cdot p_3)(p_4\cdot p_5)\over 
	(2 p_2\cdot p_5 + m_W^2)^2(p_1\cdot p_3 +\delta^2)^2 }
\ee
in the limit of massless particles, where
\be
	\delta = (E_1 - E_3)\Gamma_\nu
\ee
for the diagram with virtual $\nu$.
The  parton-level cross section 
in terms of the 
three-body phase space can be written as
\be
	\sigma ={1\over 256\,\pi^4\,\hat s}\int_0^{\sqrt{\hat
s}/2}\!\!\!\!\!\!\!\!  dE_3 \int_{\sqrt{\hat s}/2
-E_3}^{\sqrt{\hat s}/2}
\!\!\!\!\!\!\!\! dE_4\,\int_0^{2\pi}\!\!\!\! d\phi\,\int_{-1}^1
\!\!\!\! d\cos\alpha\,
 \langle|{\cal M}|^2\rangle
	\ee
where we choose coordinates such that (in the center-of-mass frame)
\bea
	\vec p_1 &=& E_1(\sin\alpha\cos\phi,
	\sin\alpha\sin\phi,\cos\alpha)\nn\\
	\vec p_2 &=& -\vec p_1\nn\\
	\vec p_3 &=& E_3\,(0,0,1)\nn\\
	\vec p_4 &=& E_4\, (\sin\gamma,0,\cos\gamma)\nn\\
	\vec p_5 &=& -\vec p_3 - \vec p_4
\eea
with $E_1 = E_2 = \sqrt{\hat s}/2$ and 
\[
	\cos\gamma = 1 + {\hat s - 2\sqrt{\hat s} (E_3 + E_4)\over 2
E_3 E_4}
\] 
being fixed by energy-momentum conservation, since $E_5 = \sqrt{\hat
s} - E_3 - E_4 = 
|\vec p_5| = |\vec p_3 + \vec
p_4|$.  Here $\hat s$ is the partonic center-of-mass energy squared,
\[
	\hat s = xs =2 x\, E_\chi\, m_N
\]
where $x$ is the quark momentum fraction and $m_N$ the nucleon mass.
Only for $x$ close to $x_{\rm min} = m_\phi^2/s$ does it matter that
the external particles are massive; otherwise the masses can be
neglected since they are $\ll 1$\,EeV, the energy scale of interest.

The momentum squared in the $\nu_\tau$ propagator is \nobreak{$2p_1\cdot
p_3$} $= 
2E_1 E_3 (1 - \cos\alpha)$, so the integral over $\cos\alpha$ is
dominated by $\cos\alpha \cong 1$ at high energies.  One finds that
\be
	I_1\equiv \int_{-1}^1 d\cos\alpha\, {p_1\cdot p_3\over (p_1\cdot p_3
+\delta^2)^2}
	= {1\over \sqrt{\hat s} E_3} \ln\left(1 + {\hat s E_3^2\over 
	\delta^2}\right)
\ee
where $\delta = (\sqrt{\hat s}/2 - E_3)/\Gamma_\nu$.
Since the most sensitive $\cos\alpha$ dependence is in this factor,
we take $\cos\alpha\to 1$ in the remaining part of $|{\cal M}|^2$,
which depends upon $E_4$.  The integral over $E_4$ can be done
analytically, and in the limit $E_3\gg m_W$ it gives
\be
	I_2\equiv \int_{\sqrt{\hat s}/2-E_3}^{\sqrt{\hat s}} dE_4\,
	{(p_2\cdot p_3)(p_4\cdot p_5)\over 
	(2 p_2\cdot p_5 + m_W^2)^2} = {\sqrt{\hat s}\,E_3^2\over 2 m_W^2} \,.
\label{I2int}
\ee
The remaining integral over $E_3$ can also be done analytically,
\be
	\int_0^{\sqrt{\hat s}/2} dE_3\, I_1 I_2 = \hat s\,
	\frac{1 + \left(1+3 \hat\Gamma ^2\right) \log (\hat\Gamma^{-1} )
+\hat\Gamma ^2 + \pi  \hat\Gamma ^3}{4
   \left(\hat\Gamma ^2+1\right)^2 m_W^2}
\label{lastint}
\ee
where $\hat\Gamma = \Gamma_\nu/\sqrt{\hat s}$.

Finally one must integrate over the parton distribution functions,
\be
	\langle \sigma\rangle = \int_{x_{\rm min}}^1 dx\,
	\sum_i f_i(x,Q)\,\sigma(x)
\ee
where we evaluated them at scale $Q = m_W$ in accordance with the low virtuality
of the internal $W$ boson indicated by (\ref{I2int}).

\end{appendix}

\bibliography{ref}

\end{document}